\documentstyle[11pt,epsfig]{article}
\date{}
\begin{document}
\title{{\bf Thermostatistics with minimal length uncertainty relation}}
\author{\\{B. Vakili}$^{1,}$\thanks{email: b-vakili@iauc.ac.ir } \nonumber\
and \nonumber\ {M. A. Gorji}$^{2,}$\thanks{email: m.gorji@sbu.ac.ir}\\\\
{\small $^1${\it Department of Physics,  Chalous Branch, IAU, P.
O. Box 46615-397, Chalous, Iran}}
\\{\small $^2${\it Department of Physics, University of Mazandaran, P.O. Box 47416-95447, Babolsar, Iran}}}  \maketitle

\begin{abstract}
Existence of minimal length is suggested in any quantum theory of
gravity such as string theory, double special relativity and black
hole physics. One way to impose minimal length is deforming
Heisenberg algebra in phase space which is called Generalized
Uncertainty Principle (GUP). In this paper, we develop statistical
mechanics in GUP framework. Our method is quite general and does
not need to fix the generalized coordinates and momenta. We define
general transformation in phase space which transforms usual
Heisenberg algebra to a deformed one. In this method, quantum
gravity effects only acts on the structure of phase space and we
relate these effects to the density of states. We find an
interesting phenomenon in Maxwell-Boltzmann statistics which has
not a classical analogy. We show that there is an upper bound for
the number of excited particles in the limit of high temperature
which implies to the condensation. Also we study modification of
Bose-Einstein condensation and the completely degenerate gas.\vspace{5mm}\noindent\vspace{0.8mm}\newline

\end{abstract}
\section{Introduction}
In Hamiltonian viewpoint of classical mechanics, canonical
equations of motion can be represented by the Poisson brackets in
which the coordinates $x_i$ and their conjugate momenta $p_j$ obey
the Poisson algebra $\{x_i,p_j\}=\delta_{ij}$. Physically, these
relations may be interpreted as the measurement of the position
and momentum of a moving particle simultaneously. This means that
the coordinates of each point of the corresponding phase space can
be determined without any uncertainty. The transition to the
quantum mechanics is straightforward. The classical dynamical
variables should be replaced by their hermitian operators
counterparts in Hilbert space and the Poisson brackets with the
Dirac commutators. Hence, the above Poisson algebra takes the form
of the Heisenberg algebra $[\hat{x}_i,\hat{p}_j]=i\hbar \delta_{ij}$. A
consequence of such a algebra is that it divides the phase space
to pixels proportional to the Planck constant and thus, the
position and the momentum of particles can not be determined
simultaneously. This is what we know as the uncertainty principle
in the usual quantum mechanics. On the other hand, one of the most
important predictions of the theories which deal with quantum
gravity is that there exists a minimal length below which no other
length can be observed \cite{AAA}. One of the interesting features
of the existence of such a minimum length is the modification it
makes to the standard commutation relation between position and
momentum in ordinary quantum mechanics which is called the
generalized uncertainty principle (GUP) \cite{BBB}. In one
dimension the simplest form of such relations can be written as
\begin{eqnarray}\label{HUP}
\bigtriangleup X \bigtriangleup P\geq \frac{\hbar}{2}\left(1+\beta
(\bigtriangleup P)^2+\gamma\right),
\end{eqnarray}where $\beta$ and $\gamma$ are
positive and independent of $\bigtriangleup X$ and $\bigtriangleup
P$, but may in general depend on the expectation values $<X>$ and
$<P>$. The usual Heisenberg uncertainty relation can be recovered
in the limit $\beta=\gamma=0$. It is easy to see that this
equation implies a minimum position uncertainty of
$(\bigtriangleup X)_{min}=\hbar \sqrt{\beta}$. For a more general
discussion on such deformed Heisenberg algebras, especially in
three dimensions, see \cite{BBB2}. Now, it is possible to realize
equation (\ref{HUP}) from the following commutation relation
between position and momentum operators

\begin{eqnarray}\label{GUP0}
\left[\hat{X}_i,\hat{P}_j\right]=\,i\hbar(1+
\beta \textbf{P}^{2})\,\delta_{ij},
\end{eqnarray}where ${\bf P}^2=P_iP^i$ and we take
$\gamma=\beta <{\bf P}>^2$. Also, assuming that

\begin{equation}\label{A}
\left[\hat{P}_i,\hat{P}_j\right]=0,\end{equation}the commutation
relations for the coordinates are obtained as

\begin{equation}\label{NNCCGUP}
\left[\hat{X}_i,\hat{X}_j\right]\,=\,2i \hbar \beta\,
(\hat{P}_i \hat{X}_j-\hat{P}_j \hat{X}_i),
\end{equation}which means that in more than one dimension,
GUP naturally implies a noncommutative geometric generalization
of position space.

In a statistical mechanics point of view, the microstate of a
given classical system may be defined by $3N$ position coordinates
$X_1,...,X_{3N}$ and $3N$ momenta coordinates $P_1,...,P_{3N}$,
where $N$ is the number of particles in the system. In a geometric
picture, the set of coordinates $(X_i,P_j)$, where $i=1,...,3N$,
may be considered as a point in a $6N$ dimensional space, the
so-called phase space of the system. Since the coordinates $X_i$
and $P_i$ vary with time, the dynamics of the whole system can be
determined with the help of the Hamiltonian equations of motion
for each of these coordinates, that is, $\dot{X_i}=\{X_i,H\}$ and
$\dot{P_i}=\{P_i,H\}$, where $H(X_i,P_j)$ is the Hamiltonian of
the system. Therefore, the dynamical behavior of the system may be
viewed as a continuous trajectory of the phase point $(X_i,P_i)$
in the phase space. On the other hand, it is well known that the
correct description of such a system and its evolution requires
quantum mechanics. In this case, the quantum particle has no
well-defined trajectory in the phase space. This is because that
at a given time $t$, there is no longer an exact position, but a
typical extent of the wave function inside which this particle can
be found. Another way of understanding this issue is to state the
Heisenberg uncertainty principle which is a result of the
commutation relations between position and momentum operators. In
this sense, if the position coordinate $X_i$ is known up to
$\bigtriangleup X_i$, then its conjugate momentum $P_i$ cannot be
determined with an accuracy better than $\bigtriangleup P_i$, such
that $\bigtriangleup X_i \bigtriangleup P_i \geq \hbar/2$. The
phase space is therefore divided into cells of area of the order
of $h^{3N}$. The accuracy on the phase space trajectory is thus
limited and this produces a discretization of the phase space.
Now, what happened if one takes into account the GUP
considerations instead of the ordinary uncertainty principle? The
motivation of this question is that the GUP scheme relies on a
modification of the canonical prescriptions and, in this respect,
can be reliably applied to any dynamical system. Over the past few
years, a number of works have been done, in the area of
statistical mechanics in the GUP framework. For instance, the
thermodynamics of the ideal gas and ultra-relativistic gas in
micro-canonical ensemble in the GUP framework are studied in
\cite{CCC}. For harmonic oscillators and ideal gases in canonical
ensembles see \cite{CCC1}. The deformed density matrix is studied
in \cite{DDD1} and modified uncertainty relation for inverse
temperature and internal energy is addressed in \cite{DDD}. Black
body radiation with minimal length effects is considered in
\cite{FFF1}. For thermodynamics in doubly special relativity
theory which also predicts a minimal length, see\cite{FFF}.
Thermodynamics in the framework of non-commutative Heisenberg
algebra (q-deformed) is investigated in \cite{EEE}.

In this paper, our aim is to study the aspects related to the
application of GUP in the framework of classical and quantum
statistical mechanics. Our study covers the thermodynamics of the
fermionic and bosonic systems when their underlying canonical
structures are of the type of GUP-deformed. We deal with the
density of deformed phase space and study the effects of existence
of minimal length on the density of states. We show that there
exist a general transformation which transforms the usual
Heisenberg algebra to a deformed one and after we convince
ourselves to the existence of such a transformation, we obtain a
density of state with the help of the Jacobian of the
transformation. Then, we study the thermodynamics of the ideal gas
and extreme relativistic gas with the help of density of deformed
phase space. The Bose-Einstein condensation and the completely
degenerate gas are the case studies which we will deal with them
in the GUP framework. Also, we find an interesting phenomenon in
the Maxwell-Boltzmann statistics which we interpret it as the
$\textit{gravitational condensation}$ and we show that it occurs
for the ideal gas and the extreme relativistic gas in the same
manner.
\section{Statistical mechanics}
Usually introducing the GUP version of a quantum theory is done by
replacing the ordinary commutation relations by their generalized
counterparts (\ref{GUP0}) and (\ref{NNCCGUP}). However, in what
follows we act differently and define a transformation in the
phase space which transforms the usual Heisenberg algebra to a
deformed one. In this picture, GUP effects only change the density
of the phase space of a statistical system. As we will see the
volume of phase space becomes larger while its degrees of freedom
decreases.

The microstates of any physical system are determined with quantum
mechanics and its energy levels should be obtained from the
Schr\"{o}dinger equation. In GUP framework, the Schr\"{o}dinger
equation becomes a non-linear or higher order differential
equation and it is not easy to solve it. For example, for the wave
function and energy spectrum of harmonic oscillator see \cite{BBB}
and \cite{BBB3}. Particular non-linear Schr\"{o}dinger equation in
GUP framework is proposed in \cite{BBB4} and solved for quantum
bouncer in \cite{BBB5}. For the higher order modified
schr\"{o}dinger equation for quantum mechanical systems, see
\cite{BBB6}. Of course, solving the higher order differential
equation has some dramatic consequences because of the
mathematical difficulty due to non-linearity and initial values
for higher order differential equations. In our method it is not
necessary to solve the modified Schr\"{o}dinger equation, instead,
we work in semi-classical approximation and we use the classical
Hamiltonian in deformed phase space. In quantum picture, energy of
a given macrostate is a summation of the energies of all
corresponding microstates. On the other hand, energy of the
macrostate is given by the Hamiltonian which is a continuous
function of the phase space variables. It is customary to
approximate the summation over the energies of the microstates by
the integral over all phase space variables. As we have mentioned,
the phase space is constructed by the fundamental cells as order
of Planck constant $\hbar$, \cite{MSM1}. We show that the
fundamental cell becomes larger and also has a momentum dependence
as $\hbar_{_{GUP}} =\, \hbar\,(1\,+\,\beta\, P^2)$. Then, we can
approximate summation over energies of microstates by integral
over all deformed phase space with spotting lattice structure of
phase space. Let us now explain our approach to transition from
ordinary phase space to deformed phase space and also show the
momentum dependence of the $\hbar_{_{GUP}}$. In ordinary quantum
mechanics, the coordinates $x_{i}$ and momenta $p_{i}$ satisfy
Heisenberg algebra $[\hat{x}_{i}\,,\,\hat{p}_{j}]=\,
i\hbar\,\delta_{ij}\,$ where its classical counterpart is the
Poisson algebra $\{x_{i}\,,\,p_{j}\}=\,\delta_{ij}\,$. In the GUP
framework, these relations take the form

\begin{eqnarray}\label{GUP1}
\{X_{i},X_{j}\}\,=\, 2\beta(P_{i} X_{j} - P_{j}
X_{i}),
\end{eqnarray}
\begin{eqnarray}\label{GUP2}
\{X_{i},P_{j}\}\,=\,(1+\beta P^2)\,\delta_{ij},
\end{eqnarray}
\begin{eqnarray}\label{GUP3}
\{P_{i},P_{j}\}\,= 0.
\end{eqnarray}

According to the Darboux theorem \cite{MSM2}, it is always
possible to find canonically conjugate variables $X_{i}(x,p)$ and
$P_{i}(x,p)$ such that they satisfy relation (\ref{GUP2}) and
(\ref{GUP3}). However, we would like to find variables $X_{i}$ and
$P_{i}$ which satisfy also relation (\ref{GUP1}) simultaneously.
In Appendix A, we convince ourselves that it is always possible to
find generalized variables $X_{i}$ and $P_{i}$ which satisfy
relations (\ref{GUP1}), (\ref{GUP2}) and (\ref{GUP3})
simultaneously. This possibility allows us to define general
transformation

\begin{eqnarray}\label{transformation}
\big(x_{i},p_{i}\big)\, \rightarrow\,
\big(X_{i}(x,p)\,,\,P_{i}(x,p)\big)\,,
\end{eqnarray}
which transforms the usual Heisenberg algebra to a deformed one.
Now, we only should have the Jacobian of this transformation for
our study. The Jacobian $\textit{J}$ of the transformation can be
expressed in terms of poisson brackets in
$2\mathcal{N}$-dimensional phase space (\,\cite{CCC1}, see also
Appendix A),

\begin{equation}\label{Jacobian}
\textit{J}\,=\,\frac{\,\partial(X_{i}\,,P_{i})}{
\partial(x_{i}\,,p_{i})}\,=\,\prod_{i=1}^{^{\mathcal{N}}}
\{X_{i}\,,P_{i} \} =\, (1+ \beta P^2)^{^{\mathcal{N}}}.
\end{equation}
In Appendix A we show that the relation (\ref{GUP1}) does not
contribute in Jacobian, thus we miss the effects of
non-commutativity in our study. Having the Jacobian of the
transformation at hand, we can define the state density
$a(\varepsilon)\, d \varepsilon$ in GUP framework by which we may
approximate the summation of energy spectrums with an integral
over deformed phase space. In the limit of large volume $V$, we
can replace summation over energy spectrums with an integral as
$\sum \rightarrow \int\frac{d\omega}{\omega_{0}}$, where
$d\omega={d^{^{\mathcal{N}}} x\,\, d^{^{\mathcal{N}}} p}$ and
${\mathcal{N}}$ is the number of degrees of freedom. Here,
$\omega_{0}$ is a fundamental volume and is given by
$\omega_{0}=h^{\mathcal{N}}$, where $h$ is the Planck constant. We
define ordinary density of state as $\sum \rightarrow \int
a_{0}(\varepsilon) d\varepsilon$. Now, we can define the density
of states in GUP framework with the help of transformation
(\ref{transformation}). In terms of the generalized variables
$X_{i}$ and $P_{i}$, the density of states for the deformed phase
space becomes

\begin{eqnarray}\label{DOSGUP}
a(\varepsilon)\, d \varepsilon \,=\,
\frac{1}{h^{^{\mathcal{N}}}}\frac{d^{^{\mathcal{N}}}
X\,\, d^{^{\mathcal{N}}}P}{\textit{J}} =\,
\frac{d^{^{\mathcal{N}}} X\,\,d^{^{\mathcal{N}}} P}
{[\,h(\,1\,+\, \beta P^2\,)]^{^{\mathcal{N}}}}\, .
\end{eqnarray}
From the above relation, it is clear that the fundamental volume
element of the corresponding phase space takes the form
$\hbar(1+\beta P^2)$ which is larger in comparison with the
ordinary case. Thus, the number of accessible microstates for the
system will decrease in GUP framework. Relation (\ref{DOSGUP})
plays an important role in our approach and we formulate the
ensemble theory in GUP framework with the help of it in this
section.

\subsection{Micro-canonical ensemble}
In equilibrium statistical mechanics, a system is supposed to be
in the thermal equilibrium and different ensembles have different
density of microstates in phase space. In pervious section we
obtain density of the deformed phase space, then it is easy to
generalize ensemble density for any ensemble in GUP framework. In
micro-canonical, the energy of a given macrostate remains constant
along the thermodynamical processes. This condition imposes the
energy constraint to the system. Under this condition the
probability of all microstates is same and density of state is
constant. We assume that the density of states to be
$\rho(x,p)^{mce} \propto \delta\big(H(x,p)-E\big)$ (see
\cite{MSM3}), where $\delta$ is the Dirac delta function, which
clearly satisfies two conditions, the same probability for all
microstates and energy constraint. Now we define the number of
microstates for $N$ particles in GUP framework as

\begin{equation}\label{MSOMCE}
\Sigma(\beta) = {\frac{1}{h^{^{3N}}}\, \int \int\,
\frac{d^{^{3N}}X d^{^{3N}}P} {{(\,1\,+\, \beta P^2)^{^{3N}}}}\,\,
\,\prod_{i=1}^{^{N}}\,\delta\big(H_{_{i}}(X,P) - E'_{i}\big)\,},
\end{equation}
where $E' =\, E'(\beta,\,E)$ is energy of each particle in GUP
framework \footnote{ We work in unit where $k_{_{b}}=\,1\,=\,c\,$,
where $k_{_{b}}$ is the Boltzmann constant and $c$ is the speed of
light.}. It is clear that the number of accessible microstates
decrease because of the GUP correction in denominator which may be
interpreted as a lost of information (see \cite{CCC0} and
\cite{CCC}). Now, the entropy of the system can be easily obtained
from its standard definition as

\begin{eqnarray}\label{ENTRPYMCE}
{\mathcal{S}}=\, \ln \Sigma(\beta).
\end{eqnarray}
There is an interesting case, because of the decreasing number of
microstates, entropy decreases and tends to zero in high energy
regime (high temperature limit). In the very early theory of
universe zero entropy coincides with preliminary singularity of
universe (highest level of energy) in which there is no physical
information as we expect from the big-bang theory. Thermodynamics
of micro-canonical ensemble can be obtained from entropy
(\ref{ENTRPYMCE}).

\subsection{Canonical ensemble}
In canonical ensemble the energy of a system is variable and
system can exchange energy. It means that the energy of the
microstates determine the probability of them in phase space.
Then, the density of microstates should be a function of energy
and we know it is proportional to Boltzmann factor:
$\rho(x,p)^{ce} \propto \exp\big(-\frac{H(x,p)}{T}\big)\,$. The
partition function of the system is given by

\begin{equation}\label{PF}
Z=\, \sum_{\varepsilon}\,\exp(-\frac{\varepsilon}{T}),
\end{equation}
where $\varepsilon$ is the energy of the microstates and is a
solution of Schr\"{o}dinger equation. In fact we should solve
modified Schr\"{o}dinger equation in GUP framework and substitute
energy of microstates in (\ref{PF}). However, in our approach we
approximate summation over energy levels by an integral with the
help of relation (\ref{DOSGUP}). Therefore, for a system with $N$
particles we get

\begin{equation}\label{CPF3}
Z_{_{N}}(T,V,\beta) =\, \frac{1}{h^{^{3N}}}\, \int\int
\frac{d^{^{3N}}X\, d^{^{3N}}P}{(\,1\,+\, \beta P^2)^{^{3N}}}\,\,
\exp\big(-\frac{H(X,P)}{T}\big).
\end{equation}
Now, all of the thermodynamical quantities in GUP framework can be
obtain from the partition function \cite{CCC1}.

\subsection{Grand canonical ensemble}

In grand canonical ensemble, in addition of its energy, the system
can be exchange its particles. Then we expect the probability of
microstates depend to the number of particles and energy. As same
as the pervious section the grand partition function in GUP
framework is

\begin{eqnarray}\label{GPF}
\Xi(T,V,N,\beta) =\, \sum_{_{N=0}}^{^{\infty}} z^{^{N}}
Z_{_{N}}(T,V,\beta),
\end{eqnarray}
where $z=e^{\mu/T}$ is the fugacity of the system and $\mu$ is the
chemical potential. Again all the thermodynamical quantities can
be obtained from grand canonical partition function. On the other
hand, one may work with $\textit{mean occupation number}$,
$<n_{{\varepsilon}}>$ instead of grand partition function
(\ref{GPF}) which can be defined as

\begin{equation}\label{NDND}
<n_{{\varepsilon}}> \,=\, \frac{1}{z^{^{-1}} e^{{\varepsilon
/T}}\,+\, \epsilon},
\end{equation}
where $\epsilon$ is a constant with the values $\epsilon=1$ (for
Fermi-Dirac statistics) and $\epsilon=-1$ (for Bose-Einstein
statistics). The particle density and pressure can be obtain from
$<n_{{\varepsilon}}>$ by definitions

\begin{eqnarray}\label{NNGUP1}
N =\, \sum_{{\varepsilon}} <n_{{\varepsilon}}>\,\,\, ,
\end{eqnarray}

\begin{equation}\label{PPGUP1}
\frac{{\mathcal{P}}V}{T}\, =\,\frac{1}{\epsilon}\,
\sum_{{\varepsilon}}\, \ln(1+ \epsilon z e^{{-{\varepsilon /
T}}}).
\end{equation}
In the limit of Large volume, we can replace the above summation
by an integral over all deformed phase space with the help of
density of state in GUP framework (\ref{DOSGUP}). Fortunately,
there is a closed form for the solution in Maxwell-Boltzmann
statistics correspond to the limit $z\rightarrow 0$. In this limit
we have a classical limit of minimal length and we only neglect
the quantum effects, but the GUP effects still exist. On the other
hand, in fully quantum area ($z \rightarrow 1$), we expect to find
the purely quantum phenomenon such as Bose-Einstein condensation.
In these limit, there is an analytical solution for any value of
$z$ in term of Fermi-Dirac and Bose-Einstein functions (see
Appendix B). We study some consequences of minimal length for two
gaseous systems, $\textit{Ideal gas}$ and $\textit{Extreme
relativistic gas}$ in next section and consider thermodynamics of
them in Maxwell-Boltzmann, Fermi-Dirac and Bose-Einstein
statistics.

\section{Maxwell-Boltzmann statistics}
To study the effects of minimal length in thermodynamics, we can
exert GUP condition with the help of relation (\ref{DOSGUP}), but
we should note when we approximate summation in relation
(\ref{NNGUP1}) by integral, the contribution of the first term
which corresponds to the ground state energy of particles
($\varepsilon=0$) missed. Thus, we decompose the first term as

\begin{eqnarray}\label{NNGUP2}
N_{e} = N -\, N_{0} = \int\, <n_{{\varepsilon}}> \,
a(\varepsilon)\, d\varepsilon,
\end{eqnarray}
where $N_{e}$ and $N_{0}$ denote the number of particles in the
excited states ground state respectively. The value of $N_{0}$
which corresponds to $\frac{z}{1-z}$ is negligible for classical
limit $z \rightarrow 0$ and becomes important in the limit of
fully quantum area $z \rightarrow 1$. However, we keep this term
in both cases. In Maxwell-Boltzmann statistics, the quantum
effects are negligible and the type of particles (fermion or
boson) is not important. Then, our start point is relations
(\ref{NNGUP1}) and (\ref{PPGUP1}) according to which we have

\begin{eqnarray}\label{NNGUP1CLS}
N^{cl} =\lim_{z\rightarrow\,0} \Bigg(\sum_{{\varepsilon}}
<n_{{\varepsilon}}> \Bigg)= z\, \sum_{{\varepsilon}}
e^{{-{\varepsilon /T}}}\,,
\end{eqnarray}
and
\begin{eqnarray}\label{PPGUP1CLS}
\frac{{{\mathcal{P}}}^{cl}V}{T}\, =\lim_{z\rightarrow\,0}
\Bigg(\frac{1}{\epsilon}\, \sum_{{\varepsilon}}\, \ln(1+ \epsilon
z e^{{-{\varepsilon / T}}})\Bigg)\, = z\, \sum_{{\varepsilon}}
e^{{-{\varepsilon /T}}},
\end{eqnarray}
which resemble Maxwell-Boltzmann distribution as we expected for
classical systems. It is seen that both of the above relations
have the same value in this limit. In fact, they are a result of
equation of state in classical statistical mechanics and by
combining them, we get

\begin{eqnarray}\label{EOSGUP1}
{{\mathcal{P}}^{cl}} V\,=\, N^{cl} T,
\end{eqnarray}
which is a familiar classical equation of state. Thus the form of
equation of states preserve for statistical systems in GUP theory.
It is important to realize that the pressure
${{\mathcal{P}}^{cl}}$ and $N^{cl}$ are in GUP framework and have
different forms in comparison with their ordinary cases. Now, we
calculate ${{\mathcal{P}}^{cl}}$ and $N^{cl}$ for two gaseous
systems, the ideal gas and extreme relativistic gas.

\subsection{Ideal gas}
We consider a gaseous system consisting of $N$ non-interacting
particles confined in volume $V$ which can sharing energy
together. The energy momentum relation for ideal gas is given by
$\varepsilon=\, \frac{P^2}{2m}$, where $m$ is the mass of each
particle. Then for sufficient large volume the summation in
relation (\ref{NNGUP1CLS}) can be replaced by an integral over all
phase space with the help of density of states in GUP framework
(\ref{DOSGUP}). By substituting (\ref{DOSGUP}) in
(\ref{NNGUP1CLS}), we have

\begin{eqnarray}\label{NNGUP1CLSIG}
N^{cl}_{e}\,\simeq\,z\,\int_{_{V}}\int_{_{0}}^{^{\infty}}
\frac{d^3 X d^3 P}{[h(1+\,\beta P^2)]^3}\,
e^{{-{\frac{P^2}{2mT}}}}.
\end{eqnarray}
The integral over coordinates gives the three dimensional volume
element $V$ and we get

\begin{eqnarray}\label{NNGUP1CLSIG}
N^{cl}_{e} =\frac{V z}{4 h^3}\big(
\frac{\pi}{\beta}\big)^{3/2}\Biggr[
\frac{\sqrt{2\gamma}(1+\gamma)}{{\gamma}^2}\,+\,
\sqrt{\pi}\, e^{1/\gamma}
\frac{({\gamma}^2-2\gamma-1)}{{\gamma}^2}
\bigg(1-\,\mbox{erf}(\frac{1}{\sqrt{\gamma}})
\bigg)
\Biggr]\,,
\end{eqnarray}
where $\gamma= \beta mT$ and $\mbox{erf}$ is the error function.
Here $N^{cl}_{e}$ denotes the number of excited particles for
ideal gas in gravitational statistics. We can remove the gravity
effects in the limit $\beta\rightarrow 0$

\begin{eqnarray}\label{NNGUP1CLSIG1}
\lim_{\beta\rightarrow\,0}N^{cl}_{e}=\frac{V}{{\lambda}^3}
\,z,
\end{eqnarray}
where $\lambda = \frac{h}{\sqrt{2 \pi m T}}$ is the mean thermal
wavelength, in agreement with usual statistical physics. Relation
(\ref{NNGUP1CLSIG}) has a interesting consequence in the limit of
$T\rightarrow\infty$ as

\begin{eqnarray}\label{NNGUP1CLSIG3}
\lim_{T\rightarrow\,\infty}N^{cl}_{e}=\frac{V z}{4 h^3}
\frac{{\pi}^2}{{\beta}^{3/2}}.
\end{eqnarray}
The above relation shows that there is an upper bound for the
number of excited particles in the limit of very high temperature
which is a purely quantum gravity effect and does not have a
classical analogy. More precise, chemical potential is the mean
energy which one particle need to apply to statistical system,
then if we increase the number of particles, the energy of the
system becomes large and larger. In ordinary Maxwell-Boltzmann
statistics, the particles access to the upper energy levels in the
limit of high temperature and there is not any limitation for
them. However in GUP theory, there is a highest energy level for
systems. Then, after the energy of the system reaches to the
highest level, if we increase the number of particles, these
should be lie in ground state and we have a condensation for
classical ideal gas. Such a condensation is different from
ordinary ones in Maxwell-Boltzmann statistics which is occurs in
the limit of low temperature. By having ${N^{cl}_{e}}$ from
relation (\ref{NNGUP1CLSIG}), the pressure of the system can be
obtained from relation (\ref{EOSGUP1}) as

\begin{eqnarray}\label{PPGUPCLSFNL}
{{\mathcal{P}}^{cl}}=\frac{z\,{\pi}^{3/2}}{4m h^3}
\frac{1}{{\beta}^{5/2}}\,\Biggr[\frac{\sqrt{2\gamma}
(1+\gamma)}{{\gamma}}\,+\,\sqrt{\pi}\,e^{1/\gamma}
\frac{({\gamma}^2-2\gamma-1)}{\gamma}\bigg(1-\,
\mbox{erf}(\frac{1}{\sqrt{\gamma}})\bigg)\Biggr]\,.
\end{eqnarray}
Also, the energy of the system can be obtained by definition

\begin{eqnarray}\label{ENRGYGUP}
\frac{{U}}{V}=\,T^2\,\Biggr[\,\frac{\partial}{\partial T}
\Big(\,\,\frac{{\mathcal{P}}}{T}\,\,\Big)\Biggr]_{_{z}},
\end{eqnarray}
from which we get

\begin{eqnarray}\label{ENRGYGUPCLSIG}
U^{cl}=\frac{z V\,{\pi}^{3/2}}{8m h^3}\frac{1}{{\beta}^{5/2}
\,{\gamma}^{2}}\,\Biggr[\,2\sqrt{\pi}\,e^{1/\gamma}
\sqrt{\gamma}\,(\gamma^2+ 4 \gamma+1)\mbox{erfc}(\frac{1}
{\sqrt{\gamma}})\,\nonumber\\ -\gamma\,\Big(\,(\sqrt{2}-2)
\,\gamma^2+(3\sqrt{2}+4)\gamma+2\,\Big)\Biggr]\,,
\end{eqnarray}
where $\mbox{erfc}$ is the complementary error function. The
pressure and energy of the classical ideal gas in purely
classical limit (quantum effects are negligible) do not show the unusual
behavior and have ordinary interpretation.

\subsection{Extreme relativistic gas}

As same as pervious section we consider $N$ non-interacting
particles which confined in volume $V$, but here the energy
momentum relation for extreme relativistic gas is given by
$\varepsilon=\,P\,$. For large volume, we can replace summation in
(\ref{NNGUP1CLS}) with the help of density of state (\ref{DOSGUP})
and we have

\begin{eqnarray}\label{NNGUP1CLERG1}
N^{cl}_{e}=\frac{V\,z}{h^3}\,\frac{\sqrt{\pi}}
{{\beta}^{3/2}}\,\,G^{^{3\,\,1}}_{_{0\,\,0}}
\left(\frac{1}{4\beta T^2}\Bigg|
\begin{array}{ccc}
\frac{-1}{2}  &  &  \\
0 & \frac{1}{2} & \frac{3}{2} \\
\end{array}
\right),
\end{eqnarray}
where $G$ is the Meijer function. Like the ideal gas, we find a
condensation due to the GUP theory for extreme relativistic gas
in the limit of high temperature

\begin{eqnarray}\label{NNGUP1CLERG3}
\lim_{T\rightarrow\,\infty}N^{cl}_{e}=\frac{V z}
{4 h^3}\frac{{\pi}^2}{{\beta}^{3/2}}.
\end{eqnarray}
Thus, again there is an upper limit for the number of particles in
the limit of high temperature $T\rightarrow\,\infty$. The maximum
value of the number of particles for extreme relativistic gas is
as same as the ideal gas (see relation (\ref{NNGUP1CLSIG3})). The
pressure can be obtained by substituting (\ref{NNGUP1CLERG1}) in
(\ref{EOSGUP1}). For energy $U^{cl}$ of the system we have

\begin{eqnarray}\label{ENRGYGUP1CLS}
U^{cl}\,=\frac{\sqrt{\pi}}{2 h^3\,{\beta}^{5/2}}\,
\frac{z V}{T} G^{^{3\,\,1}}_{_{0\,\,0}}
\left(\frac{1}{4\beta T^2}\Bigg|
\begin{array}{ccc}
\frac{-3}{2} &   &  \\
\frac{-1}{2} & 0 & \frac{1}{2} \\
\end{array}
\right).
\end{eqnarray}
In the limit of high temperature $T\,\rightarrow\,\infty$, for
the above relation we have

\begin{eqnarray}\label{ENRGYGUP1CLS2}
U^{cl}\,=\frac{V z}{h^3}\frac{\pi}{{\beta}^{2}}.
\end{eqnarray}
Thus, there is an upper bound for energy of the extreme
relativistic gases in GUP framework which is as order of Planck
energy $U^{cl}\,\sim\,E_{_{pl}}$ (note
$\beta\sim\frac{1}{E_{_{pl}}^2}$ in our unit).
\section{Fermi-Dirac and Bose-Einstein statistics}

In this limit the type of particles, which may be fermion or
boson, plays an important role, according to which there are two
types of statistics the so-called Fermi-Dirac and Bose-Einstein
for fermions and bosons respectively. To consider the
thermodynamics of fermions and bosons, we start again with
relations (\ref{NNGUP1}) and (\ref{PPGUP1}). We replace summation
in these relations by integral over all deformed phase space with
the help of relation (\ref{DOSGUP}). As we mentioned, these
integrals do not have a closed form solution, but fortunately they
have an analytical solution in the limit of $\beta\rightarrow 0$
(which is a good approximation, because of the relation $\beta
\sim \frac{1}{{E_{_{pl}}}^2}$) and we calculate them in terms of
Fermi-Dirac and Bose-Einstein functions for general power law
energy-momentum relation $\varepsilon=\,\eta P^{\alpha}$ in
Appendix B. The number of excited particles and pressure for
fermions and bosons (see relations (\ref{NUMBER2}) and
(\ref{GUPFFFF})) are given by

\begin{eqnarray}\label{NUMBER00}
N_{e}=\frac{4\pi g}{\alpha h^3}\,(T/\eta)^{^{3/\alpha}}\,
\Gamma(3/\alpha)\,\,h_{_{3/\alpha}}(z)\,\Biggr[\,1\,-\,3
\beta\,(T/\eta)^{^{2/\alpha}}\,\frac{\Gamma(5/\alpha)\,\,
h_{_{5/\alpha}}(z)}{\Gamma(3/\alpha)\,\,h_{_{3/\alpha}}(z)}
\nonumber\\ \,+\,6{\beta}^2\,(T/\eta)^{^{4/\alpha}}\,\frac{
\Gamma(7/\alpha)\,\,h_{_{7/\alpha}}(z)}{\Gamma(3/\alpha)\,
\,h_{_{3/\alpha}}(z)}\,+\,...\,\Biggr]\,,
\end{eqnarray}
and

\begin{eqnarray}\label{PPGUP00}
\frac{{\mathcal{P}}}{T}=\frac{4\pi g}{\alpha h^3}\,
(T/\eta)^{^{3/\alpha}}\,\Gamma(3/\alpha)\,\,
h_{_{\frac{3}{\alpha}+1}}(z)\,\,\Biggr[\,\,1\,-\,\,
3\,\beta\,(T/\eta)^{^{2/\alpha}}\,\,\frac{\Gamma(5/\alpha)
\,h_{_{\frac{5}{\alpha}+1}}(z)}{\Gamma(3/\alpha)\,
h_{_{\frac{3}{\alpha}+1}}(z)}\,\,\nonumber\\+\,6\,
{\beta}^2\,(T/\eta)^{^{4/\alpha}}\,\frac{\Gamma(7/\alpha)
\,h_{_{\frac{7}{\alpha}+1}}(z)}{\Gamma(3/\alpha)\,
h_{_{\frac{3}{\alpha}+1}}(z)}+\,\,...\Biggr]\,,
\end{eqnarray}
where $g$ is a weight factor showing the internal degree of
freedom which for bosons and spin-$1/2$ particles is equal $1$ and
$2$ respectively. Here, $h_{_{\nu}}(z)$ is given by relation
(\ref{HFDBE}) and reduces to Fermi-Dirac and Bose-Einstein
functions according to the type of particles. Now, we can obtain
energy of the systems by using of the definition (\ref{ENRGYGUP})
as

\begin{eqnarray}\label{ENRGYGUP2}
U=\frac{3VT}{\alpha}\frac{4\pi g}{\alpha h^3}\,
(T/\eta)^{^{3/\alpha}}\,\Gamma(3/\alpha)\,\,
h_{_{\frac{3}{\alpha}+1}}(z)\,\,\Biggr[\,\,1\,-\,\,5\,
\beta\,(T/\eta)^{^{2/\alpha}}\,\,\frac{\Gamma(5/\alpha)
\,h_{_{\frac{5}{\alpha}+1}}(z)}{\Gamma(3/ \alpha)\,
h_{_{\frac{3}{\alpha}+1}}(z)}\,\,\nonumber\\+\,14\,
{\beta}^2\,(T/\eta)^{^{4/\alpha}}\,\frac{\Gamma(7/\alpha)
\,h_{_{\frac{7}{\alpha}+1}}(z)}{\Gamma(3/\alpha)\,
h_{_{\frac{3}{\alpha}+1}}(z)}+\,\,...\Biggr]\,.
\end{eqnarray}
We can combine relations (\ref{PPGUP00}) and (\ref{ENRGYGUP2}) to
get equation of state for power law energy-momentum statistical
systems

\begin{eqnarray}\label{EQOSTATE00}
\frac{\mathcal{P}}{\rho}=\frac{\alpha}{3}\Biggr[\,\,1\,+\,\,2\,
\beta\,(T/\eta)^{^{2/\alpha}}\,\,\frac{\Gamma(5/\alpha)\,
h_{_{\frac{5}{\alpha}+1}}(z)}{\Gamma(3/\alpha)\,h_{_{\frac{3}
{\alpha}+1}}(z)}\,\,+\,2\,{\beta}^2\,(T/\eta)^{^{4/ \alpha}}\,
\nonumber\\ \Bigg(\,5\,\Bigg(\frac{\Gamma(5/\alpha)\,
h_{_{\frac{5}{\alpha}+1}}(z)}{\Gamma(3/\alpha)\,h_{_{\frac{3}
{\alpha}+1}}(z)}\,\Bigg)^2\,-\,4\,\frac{\Gamma(7/\alpha)\,
h_{_{\frac{7}{\alpha}+1}}(z)}{\Gamma(3/\alpha)\,h_{_{\frac{3}
{\alpha}+1}}(z)}\,\Bigg)+\,\,...\Biggr]\, ,
\end{eqnarray}
where $\rho =U/V$ is the energy density. The l.h.s. of above
equation denotes the equation of state (EoS) parameter $\omega$
which is a function of temperature in GUP framework. The first
term in the r.h.s. is an ordinary EoS parameter and the other
terms exhibit the quantum gravity effects which, in a cosmological
point of view, are very small in the late times of cosmic
evolution. However, in the planck scale they become important and
play a determinant role in early universe thermodynamics. Then, we
rewrite the above equation in the form

\begin{eqnarray}\label{EQOSTATE01}
\omega(\beta,T)=\,\omega_{0}\Biggr[\,\,1\,+\,\,2\,\beta\,
(T/\eta)^{^{2/\alpha}}\,\,\frac{\Gamma(5/\alpha)\,h_{_{\frac{5}
{\alpha}+1}}(z)}{\Gamma(3/\alpha)\,h_{_{\frac{3}{\alpha}+1}}
(z)}\,\,+\,2\,{\beta}^2\,(T/\eta)^{^{4/\alpha}}\,\nonumber\\
\Bigg(\,5\,\Bigg(\frac{\Gamma(5/\alpha)\,h_{_{\frac{5}{\alpha}+1}}
(z)}{\Gamma(3/\alpha)\,h_{_{\frac{3}{\alpha}+1}}(z)}\,\Bigg)^2\,
-\,4\,\frac{\Gamma(7/\alpha)\,h_{_{\frac{7}{\alpha}+1}}(z)}
{\Gamma(3/\alpha)\,h_{_{\frac{3}{\alpha}+1}}(z)}\,\Bigg)+\,\,...
\Biggr]\,.
\end{eqnarray}
Now, at the first, we consider above relation for non-relativistic
particles. The energy-momentum relation for the non-relativistic
particles is given by relation $\varepsilon=\frac{P^2}{2m}$, where
$m$ is the mass of the particles
($\alpha=2\,,\,\eta=\frac{1}{2m}\,$). In this case we have

\begin{eqnarray}\label{CPNR}
\omega(T)\,\simeq\,\frac{2}{3}\,\Bigr[\,1+\,5m\,\,\frac{T}
{{T_{_{Pl}}}^2}\,\,\Bigr].
\end{eqnarray}
It is clear that even at the Planck temperature $T\,=\,T_{_{Pl}}$,
the second term in the r.h.s. of this relation has a small value
because of the large value of the Planck temperature $T_{_{Pl}}
\simeq 1.221 \times 10^{19}\,\textit{Gev}$. But for relativistic
particles the energy-momentum relation is given by
$\varepsilon=\,P$ ($\alpha=\,1=\,\eta$) and relation
(\ref{EQOSTATE01}) for the bosonic gas in the limit
$z\rightarrow1$ becomes

\begin{eqnarray}\label{CPR1}
\omega(T)\,=\,\frac{1}{3}\,\Biggr[\,1+24\,\frac{\zeta(6)}
{\zeta(4)}\,\Big(\frac{T}{T_{_{Pl}}}\Big)^2\,+\,1440\,\Bigr[
\,\frac{\zeta(6)}{\zeta(4)}\,\Big(\,\frac{\zeta(6)}{\zeta(4)}
\,-2\,\frac{\zeta(8)}{\zeta(6)}\,\Big)\Bigr]\,\Big(\frac{T}
{T_{_{Pl}}}\Big)^4\,+\,...\Biggr],
\end{eqnarray}
where $\zeta(\nu)$ is the Riemmnn zeta function . Unlike the case
of non-relativistic particles, the above relation does not
converge in the limit of high temperature and all terms in the
r.h.s. become important. Since the zeta function decreases with
its argument, the coefficients of all the terms in r.h.s. expect
the second one are negative. In this sense we offer the simplest
temperature power law for the relation (\ref{CPR1}) as

\begin{eqnarray}\label{CPR2}
\omega(T)\,\equiv \,\frac{1}{3}\, \Biggr[\, 1 -\,
\sigma\,\,\Big(\frac{T}{T_{_{Pl}}}\Big)^{2q}\, \Biggr],
\end{eqnarray}
where $\sigma$ and $q \geq 2$ are some constants which may be
fixed by cosmological observation. An interesting feature of
relation (\ref{CPR2}) is when the temperature increases and
becomes larger than critical temperature $T_{c} =\,
(\frac{1}{\sigma})^{\frac{1}{2q}} T_{_{Pl}} $, for which we have a
quintessence scenario for boson gaseous system. If we set
$T_{c}=\,T_{_{Pl}}$ then relation (\ref{CPR2}) describes an
inflationary scenario. in figure $1$ we have plotted the behavior
of $\omega$ for three values of $q$ in which the dashed line shows
the ordinary case and three curves show the effects of minimal
length on cosmological parameter $\omega$ as a function of
temperature. It is clear that in the limit of low temperature all
the curves tend to ordinary case which shows that the quantum
gravity effects only become significant in high temperature i.e.
in the Planck scales.
\begin{figure}\label{CP11}
\begin{tabular}{ccc} \epsfig{figure=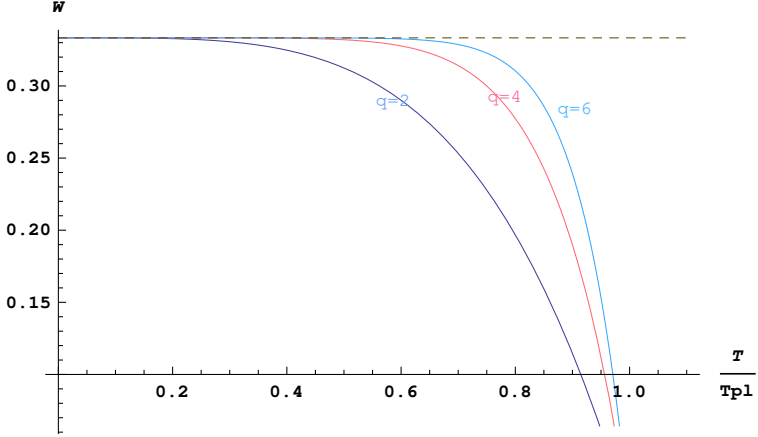,width=10cm}
\hspace{1cm}
\end{tabular}
\caption{\label{CosmologicalP1}\footnotesize The effects of the
minimal length on the EoS parameter $\omega$. As we see in the
figure, the Eos parameter is a function of temperature in the GUP
framework and get a negative value if temperature becomes more
than critical temperature $T_{c} =
(\frac{1}{\sigma})^{\frac{1}{2q}} T_{_{Pl}}$.}
\end{figure}

\subsection{Bose-Einstein condensation}

After set up the Bose-Einstein statistics, it is natural to
consider famous phenomenon Bose-Einstein condensation. The
equation (\ref{NUMBER00}) for bosons gives

\begin{eqnarray}\label{NUMBER001}
N_{e}=\frac{4\pi}{\alpha h^3}\,(T/\eta)^{^{3/\alpha}}\,
\Gamma(3/\alpha)\,\, g_{_{3/\alpha}}(z)\,\Biggr[\,1\,
-\,3 \beta\,(T/\eta)^{^{2/\alpha}}\,\frac{\Gamma(5/\alpha)
\,\,g_{_{5/\alpha}}(z)}{\Gamma(3/\alpha)\,\,g_{_{3/\alpha}}
(z)}\nonumber\\ \,+\,6{\beta}^2\,(T/\eta)^{^{4/\alpha}}\,
\frac{\Gamma(7/\alpha)\,\,g_{_{7/\alpha}}(z)}{\Gamma(3/
\alpha)\,\,g_{_{3/\alpha}}(z)}\,+\,...\,\Biggr]\,.
\end{eqnarray}
We know from the ordinary case that the fugacity of the bosonic
systems has a finite rang $\,0\,<z\,<1\,$ and Bose-Einstein
condensation occurs in the limit of $z\rightarrow 1$. Then, in
this limit the above relation becomes

\begin{eqnarray}\label{NUMBER001}
N_{e}^{^{max}} = N_{_{0e}}^{^{max}} \Biggr[ 1-\, 3 \beta (T/
\eta)^{^{2/ \alpha}} \frac{\Gamma(5/ \alpha) \zeta({5/ \alpha})
}{\Gamma(3/ \alpha) \zeta({3/ \alpha})}+\, 6 {\beta}^2 \,(T/
\eta)^{^{4/ \alpha}} \frac{\Gamma(7/ \alpha) \zeta({7/
\alpha})}{\Gamma(3/ \alpha) \zeta({3/ \alpha})} + ... \Biggr]\, ,
\end{eqnarray}
where $N_{_{0e}}^{^{max}}=\,\frac{4 \pi}{\alpha h^3}\, (T/
\eta)^{^{3/ \alpha}} \, \Gamma(3/ \alpha)\,\, \zeta({3/
\alpha})\,$ is the maximum number of excited particles in ordinary
Bose-Einstein statistics. It is clear that the remaining terms in
the r.h.s. are the quantum gravity effects and are small in low
temperature. Therefore, for the non-relativistic particles with
energy-momentum relation $\varepsilon=\,\frac{P^2}{2m}$, we get

\begin{eqnarray}\label{BECIG}
N_{e}^{^{max}} \simeq\, \frac{(2 \pi m
T)^{\frac{3}{2}}}{h^3}\,\zeta(3/2)\, \Biggr[
1-\,6m\,\frac{T}{{T_{_{Pl}}}^2}\,\, \Biggr].
\end{eqnarray}
Thus, the number of excited particles is smaller than the ordinary
case and Bose-Einstein condensation occurs sooner in comparison
with the ideal gas. However, the correction term due to the GUP
effects is very small even in Planck Temperature. For relativistic
particles with energy-momentum relation $\varepsilon=\,P$,
relation (\ref{NUMBER001}) becomes

\begin{eqnarray}\label{BECERG}
N_{e}^{^{max}} \simeq\,\frac{8 \pi}{h^3}\,T^3\,\zeta(3)\,\Biggr[
1-\,31\,\Big(\frac{T}
{T_{_{Pl}}}\Big)^2\,+\,1812\,\Big(\frac{T}{T_{_{Pl}}}\Big)^4\,+\,
...\Biggr].
\end{eqnarray}
The above relation does not converge in high temperature and all
the correction terms in the r.h.s. are important.

\subsection{Degenerate gas}

One of the interesting cases in Fermi-Dirac statistics is a
completely degenerate gas which corresponds to the limit
$T\rightarrow0$. In this limit the mean occupation number
$<n_{\varepsilon}>$ in relation (\ref{NDND}) takes a simple form
as
\newline
\newline
$\,<n_{\varepsilon}>=\left\{
\begin{array}{ll}
1  \qquad  \varepsilon < \mu_{0} \\
0  \qquad  \varepsilon > \mu_{0},
\end{array}
\right.$
\newline
\newline
where $\mu_{0}$ denotes to the chemical potential in the limit
$T\rightarrow0$. Then, mean occupation number becomes independent
of the temperature and acts as a step function. Now, the relations
(\ref{NNGUP1}) and (\ref{PPGUP1}) become very simple and have an
exact solution also in GUP framework. Then, relation
(\ref{NNGUP1}) takes the form

\begin{eqnarray}\label{CDGN1}
N\,=\,\sum_{\varepsilon}\,\simeq\, \int a(\varepsilon)
d\varepsilon.
\end{eqnarray}
Here, there is a limited value for energy which is called Fermi
energy $\varepsilon_{_{F}}$, in terms of which one can also define
the Fermi momentum $P_{_{F}}$. Upon using of relation
(\ref{DOSGUP}), we have

\begin{eqnarray}\label{CDG2}
N\,=\frac{4{\pi}g V}{h^3}\int_{0}^{P_{_{F}}}\,\frac{P^2\,dP}{(1+\beta P^2)^3}
= \frac{\pi V g}{2 h^3}\, \Biggr[\, \frac{\arctan(\sqrt\beta
P_{_{F}})}{\beta^{\frac{3}{2}}} \,+\, \frac{P_{_{F}} (\beta
{P_{_{F}}}^2 - 1)}{\beta {(1+\beta {P_{_{F}}}^2)}^2}\, \Biggr],
\end{eqnarray}
which denotes to the number of particles in completely degenerate
gaseous system with any energy-momentum relation in GUP framework.
Up to the first order of $\beta$ from the above relation one
obtains

\begin{eqnarray}\label{PFGPF1}
\frac{9}{5} \beta (P_{_{F}})^5 - ({P_{_{F}}})^3 +
{{P_{_{0}}}_{_{F}}}^3=0,
\end{eqnarray}
where ${P_{_{0}}}_{_{F}}= (\frac{3N}{4 \pi V g})^{\frac{1}{3}} h$
is the Fermi momentum in ordinary case. We get $P_{_{F}}
={P_{_{0}}}_{_{F}} $ in the limit $\beta=0$, so we expand the
above relation in the small but nonzero limit of GUP parameter
$\beta\rightarrow0$ to get

\begin{eqnarray}\label{PFGPF2}
P_{_{F}} \simeq {P_{_{0}}}_{_{F}} \Big[\,1\,+\, \frac{3}{5} \beta
{P_{_{0}}}_{_{F}}^2\, -\, \frac{24}{175} {\beta}^2
{P_{_{0}}}_{_{F}}^4\, +\, ... \Big].
\end{eqnarray}
The first term in r.h.s. of the above relation denotes the
ordinary Fermi momentum and the second term is the most important
effect of GUP on the Fermi momentum. Then, independent of that the
gaseous system being relativistic or not, minimal length effects
increase the value of the Fermi momentum and consequently increase
the Fermi energy.

\section{Conclusions}
There are some evidences which support the idea of the existence
of minimal length from quantum gravity theories. GUP theory
imposes minimal length in quantum mechanics by modifying
uncertainty relations, then the deformed Heisenberg algebra
governs the phase space instead of the usual Heisenberg algebra.
The customary procedure for exerting GUP conditions to a physical
systems is finding the generalized coordinates and momenta which
satisfy the deformed Heisenberg algebra. For the simplest choice,
the Schr\"{o}dinger equation becomes non-linear or it becomes a
higher order differential equation and it is not easy to solve it.
In this paper we have extended statistical mechanics in GUP
framework without fixing coordinates and momenta. We investigated
the possibility of existence a general transformation in phase
space which changes the usual Heisenberg algebra to a deformed
one. Such a transformation changes the density of states in phase
space. We obtained a deformed density of states for GUP conditions
by the use of the Jacobian of the transformation. In this picture,
effects of minimal length only changes the structure of phase
space, so we conclude minimal length influences the situations of
microstates in quantum scales and decreases the number of
accessible microstates of the system. The advantage of this
procedure is that we do not need to solve the modified
Schr\"{o}dinger equation, instead we can work with the classical
Hamiltonian together deformed density of state. In this method, we
do not use the wave functions, so the exact solution of modified
Schr\"{o}dinger equations reserve their importance. Also we showed
the non-commutativity of coordinates due to a GUP condition does
not contribute in the density of states till the momenta commute.
We develop the ensemble theory by this approach and we study the
Maxwell-Boltzmann, Fermi-Dirac and Bose-Einstein statistics. We
found an interesting phenomenon in Maxwell-Boltzmann statistics:
there is an upper bound for the number of excited particles for
gaseous system (ideal gas and extreme relativistic gas) in the
limit of high temperatures, which means that we have a
condensation due to a GUP theory. We also found an interesting
temperature dependent equation of state parameter $\omega$ for the
bosonic gas based on which our analysis showed these effects are
negligible for non-relativistic particles even in Planck scales.
On the other hand, these effects become very important for the
relativistic bosonic gaseous systems and lead to an inflationary
scenario in the limit of high temperatures. We showed there is a
critical temperature which can be determined in the limit of the
acceleration phase, but there is two free parameters that can be
fixed with the cosmological observations. We studied two
interesting full quantum cases, Bose-Einstein condensation and the
completely degenerate gas in the GUP framework. Our analysis shows
the number of excited particles becomes smaller and the
condensation occurs sooner. For the degenerate gas, we found a
modified closed form for the number of particles in GUP framework
and we obtained a modification to a Fermi energy and momentum.
Finally we concluded all the corrections due to a minimal length
are small in ordinary temperatures and will become important on
the high temperature limit.

\appendix

\renewcommand{\theequation}{A-\arabic{equation}}
\setcounter{equation}{0}
\section{Jacobian in terms of Poisson brackets}
In this Appendix, we would like to show that it is possible to
define general transformation in phase space which transforms
coordinates $x_{i}$ and momenta $p_{i}$ to generalized coordinates
$X_{i}$ and momenta $P_{i}$ as new phase space variable satisfy
all the relations (\ref{GUP1}), (\ref{GUP2}) and (\ref{GUP3}). It
is well-known from classical physics that the coordinates $x_{i}$
and momenta $p_{i}$ satisfy the Poisson algebra $\{ x_{i},p_{i}
\}=\delta_{ij}$. However, Darboux theorem implies that it is
always possible to find variables $X_{i}$ and $P_{i}$ as a
function of $x_{i}$ and $p_{i}$ which satisfy relation

\begin{eqnarray}\label{PBGDT}
\{X_{i}\,,\,P_{j}\}\,=\, f(\,X_{i},\,P_{j}).
\end{eqnarray}
Therefore, the relation (\ref{GUP2}) is a particular case of the
Darboux theorem and in GUP framework $f(X,P)=\,f(P)= i
\hbar(1+\,\beta{P^2}) $. On the other hand, relation (\ref{GUP3})
does not exert any constraint to our study. Thus, we can deduce
that the relations (\ref{GUP2}) and (\ref{GUP3}) are valid
simultaneously in general. But we want to have a transformation
which satisfy also relation (\ref{GUP1}). We suppose general forms
of non-commutativity as

\begin{eqnarray}\label{PBGDT2}
\{X_{i}\,,\,X_{j}\}\,=\, g(\,X_{i},\,P_{j}),
\end{eqnarray}
where $g$ is the arbitrary function of variables $X_{i}$ and
$P_{j}$. Again, it is possible to find variables $X_{i}$ and
$P_{j}$ as a function of $x_{i}$ and $p_{i}$ which satisfy
relation (\ref{PBGDT2}), but we should have  variables $X_{i}$ and
$P_{j}$ as a function of $x_{i}$ and $p_{j}$ which satisfy
relations (\ref{GUP1}), (\ref{GUP2}) and (\ref{GUP3})
simultaneously. In general, we cannot prove that it is always
possible to define general transformation which changes ordinary
variables $x_{i}$ and $p_{j}$ (which satisfy Poisson algebra) to
GUP variables $X_{i}$ and $P_{j}$ (which satisfy deformed Poisson
algebra). However, at least there is one choice $X_{i}$ and
$P_{i}$ as
\newline
\newline
$\left\{
\begin{array}{ll}
X_{i} =\, x_{i}\, \big(\, 1+\, \beta p^2 \big)\, ,\\
P_{i} =\, p_{i},
\end{array}
\right.$
\newline
\newline
which satisfy all the relations (\ref{GUP1}), (\ref{GUP2}) and
(\ref{GUP3}) simultaneously. This implies to define transformation
between usual coordinates $(x_{i}, p_{i})$ and GUP variables
$(X_{i},P_{i})$. The Jacobian of transformation can be expanded in
terms of Poisson brackets in $2\mathcal{N}$-dimensional phase
space as \cite{CCC1}

\begin{eqnarray}\label{Jacobian1}
\textit{J}\,=\,\frac{\, \partial(X_{i}\,,\,P_{i}) }{
\partial(x_{i}\,,\,p_{i}) }\,= \frac{1}{2^{^{\mathcal{N}}}
{\mathcal{N}}!}\,
\sum_{i_{1}...i_{2{\mathcal{N}}}=1}^{^{2{\mathcal{N}}}}\,
\varepsilon_{i_{1}...i_{2{\mathcal{N}}}}\, \{
J_{i_{1}},J_{i_{2}}\}\, ...\, \{
J_{i_{2{\mathcal{N}}-1}},\,J_{i_{2{\mathcal{N}}}}\}\, ,
\end{eqnarray}
where $\varepsilon$ denotes the Levi-Civita symbol and $J_{i}$
represents the phase space variables, where odd $i$ is for
coordinate $X_i$ and even $i$ is for conjugate momentum $P_i$. In
the above relation the Poisson brackets which denote only
coordinates, commutator (\ref{GUP1}), are always multiplied by
Poisson brackets that purely denote the momentum commutator,
(\ref{GUP3}). Thus the coordinates Poisson brackets do not
contribute, because the momenta commute. Therefore only Poisson
brackets that include both coordinates and momenta, (\ref{GUP2}),
contribute in relation (\ref{Jacobian1}) and the Jacobian
simplifies to (\ref{Jacobian}).

\renewcommand{\theequation}{B-\arabic{equation}}
\setcounter{equation}{0}
\section{particle density and pressure in GUP framework}

The number of excited particles in GUP framework can be obtained
from relation (\ref{NNGUP2}), by substituting mean occupation
number $<\,n_{{\varepsilon}}>$ from (\ref{NDND}), we get

\begin{eqnarray}\label{NN}
N_{e} =\, \int\,\frac{a(\varepsilon) d\varepsilon }{{z^{-1}}\,
e^{\varepsilon/T}\,+\, \epsilon},
\end{eqnarray}
where $a(\varepsilon)\, d\varepsilon$ is the density of states in
GUP framework determined by (\ref{DOSGUP}). Evaluation of this
integral depends on the energy-momentum relation which
unfortunately, there is no closed form solution for familiar power
law energy-momentum. So we expand $a(\varepsilon) d\varepsilon$
for small $\beta$. Up to the second order of $\beta$ in three
dimension (\ref{DOSGUP}) we obtain
\begin{eqnarray}\label{DOSDOSEXEX}
a(\varepsilon)\, d\varepsilon\, \simeq\, \frac{d^3 X d^3 P}{h^3}\,
\Big(\,\,1\,-\,3 \beta\, P^2\, +\, 6 {\beta}^2 \, P^4\, +\, ...
\Big)\,.
\end{eqnarray}
We consider general power law energy-momentum relation
$\varepsilon =\small{\eta} P^{^{\alpha}}$ which reduces to the
ultra-relativistic case for $\eta =1=\alpha$ and get the
non-relativistic energy-momentum relation for $\eta=1/2m$ and
$\alpha=2$. Then by substituting (\ref{DOSDOSEXEX}) in (\ref{NN})
results

\begin{eqnarray}\label{NUMBER1}
\frac{N}{V}=n=\frac{4\pi}{\alpha h^3}\,(T/\eta)^{^{3/\alpha}}
\Biggr[\,\int_{0}^{\infty}\,\frac{y^{^{\frac{3}{\alpha}\,-\,1}}
dy}{z^{-1}e^y\,+\,\epsilon}\,-\,3\beta\,(T/\eta)^{^{2/\alpha}}
\,\int_{0}^{\infty}\,\frac{y^{^{\frac{5}{\alpha}-1}}dy}
{z^{-1} e^{^{y}}\,+\,\epsilon}\,\nonumber\\+\,6{\beta}^2\,
(T/\eta)^{^{4/\alpha}}\,\int_{0}^{\infty}\,\frac{y^{^{\frac{7}
{\alpha}-1}}dy}{z^{-1}e^{^{y}}\,+\,\epsilon}\,+\,...\,\Biggr]\,,
\end{eqnarray}
where $n$ is the particle density and $y=\, {\varepsilon/T}$.
Integrals in the above relation are nothing but Fermi-dirac and
Bose-Einstein functions for $\epsilon = 1$ and $\epsilon = -1$
respectively. We define function $h_{_{\nu}}(z)$ as

\begin{eqnarray}\label{HFDBE}
h_{_{\nu}}(z)\,=\,\frac{1}{\Gamma(\nu)}\,\int_{0}^{\infty}\,
\frac{y^{^{\nu-1}}dy}{z^{-1}e^{^{y}}\,+\,\epsilon}\,=\,
\left\{
\begin{array}{ll}
f_{_{\nu}}(z)\,\, \qquad \epsilon\,=+1  \\
g_{_{\nu}}(z)\,\, \qquad \epsilon\,=-1
\end{array}
\right.
\end{eqnarray}
where $\Gamma(\nu)$ is a factorial function. Now particle density
in terms of Fermi-Dirac and Bose-Einstein functions becomes

\begin{eqnarray}\label{NUMBER2}
n=\frac{4\pi}{\alpha h^3}\,(T/\eta)^{^{3/\alpha}}\,\Gamma(3/\alpha)
\,\,h_{_{3/\alpha}}(z)\,\Biggr[\,1\,-\,3\beta\,(T/\eta)^{^{2/ \alpha}}
\,\frac{\Gamma(5/\alpha)\,\,h_{_{5/\alpha}}(z)}{\Gamma(3/ \alpha)\,\,
h_{_{3/\alpha}}(z)}\nonumber\\\,+\,6{\beta}^2\,(T/\eta)^{^{4/ \alpha}}
\,\frac{\Gamma(7/\alpha)\,\, h_{_{7/\alpha}}(z)}{\Gamma(3/\alpha)\,\,
h_{_{3/\alpha}}(z)}\,+\,...\,\Biggr]\,.
\end{eqnarray}
On the other hand, pressure of the system will be determined by
relation (\ref{PPGUP1}). Again, for large volume we can replace
summation by an integral. By using of relation (\ref{DOSDOSEXEX})
for general energy-momentum relation we get
\begin{eqnarray}\label{PPGUP2}
\frac{{\mathcal{P}}}{T}=\frac{4\pi}{\epsilon \alpha h^3
{\eta}^{^{3/\alpha}}} \Biggr[\int_{0}^{\infty}\ln(1\,+\epsilon
z e^{{-{\varepsilon/T}}}) {\varepsilon}^{^{\frac{3}{\alpha}
-1}}d\varepsilon-3\beta {\eta}^{^{-{2/\alpha}}}\int_{0}^{\infty}
\ln(1+ \epsilon z e^{{-{\varepsilon / T}}})
{\varepsilon}^{^{\frac{5}{\alpha}-1}}d\varepsilon \nonumber\\
+\,6{\beta}^2\,{\eta}^{^{-{4/\alpha}}}\int_{0}^{\infty}\,
\ln(1+\epsilon z e^{{-{\varepsilon/T}}})\,
{\varepsilon}^{^{\frac{7}{\alpha}-\,1}}d\varepsilon\,+...
\Biggr]\, .
\end{eqnarray}
Integrating by part and set $y=\varepsilon/T$ yields

\begin{eqnarray}\label{PPGUP3}
\frac{{\mathcal{P}}}{T}=\,\frac{4\pi}{3 h^3}\,(T/\eta)^{^{3/\alpha}}
\Biggr[\,\int_{0}^{\infty}\frac{y^{^{3/\alpha}}dy}{z^{-1}e^{^{y}}\,+
\,\epsilon}-\,\frac{9}{5}\beta\,(T/\eta)^{^{2/\alpha}}\int_{0}^{\infty}
\frac{y^{^{5/\alpha}}dy}{z^{-1}e^{^{y}}+\,\epsilon} \nonumber\\+\,
\frac{18}{7}{\beta}^2\,(T/\eta)^{^{4/\alpha}}\int_{0}^{\infty}
\frac{y^{^{7/\alpha}}dy}{z^{-1}e^{^{y}}+\,\epsilon}+\,...\,\Biggr]\,.
\end{eqnarray}
The above integrals are again Fermi-Dirac and Bose-Einstein
functions and we can express them in terms of $h_{_{\nu}}(z)$ as

\begin{eqnarray}\label{GUPFFFF}
\frac{{\mathcal{P}}}{T}=\frac{4\pi}{\alpha h^3}\,(T/\eta)^{^{3/\alpha}}
\,\Gamma(3/\alpha)\,\,h_{_{\frac{3}{\alpha}+1}}(z)\,\,\Biggr[\,\,1\,-\,
\,3\,\beta\,(T/\eta)^{^{2/\alpha}}\,\,\frac{\Gamma(5/\alpha)\,
h_{_{\frac{5}{\alpha}+1}}(z)}{\Gamma(3/\alpha)\,h_{_{\frac{3}{\alpha}+1}}(z)}
\,\,\nonumber\\+\,6\,{\beta}^2\,(T/\eta)^{^{4/\alpha}}\,\frac{\Gamma(7/\alpha)
\,h_{_{\frac{7}{\alpha}+1}}(z)}{\Gamma(3/\alpha)\,h_{_{\frac{3}{\alpha}+1}}(z)}
+\,\,...\Biggr]\,.
\end{eqnarray}
Relations (\ref{NUMBER2}) and (\ref{GUPFFFF}) are essential for
our study in thermodynamics of bosons and fermions in GUP
framework. \vspace{5mm}\newline \noindent {\bf
Acknowledgement}\vspace{2mm}\noindent\newline The authors would
like to thank Kourosh Nozari for a careful reading of the
manuscript and helpful comments.

\end{document}